\documentclass[aps,prl,twocolumn,showpacs,amsmath,amssymb,floatfix]{revtex4}
\usepackage{graphicx}
\usepackage{bm}
\usepackage{color}
\def\be{\begin{equation}}
\def\ee{\end{equation}}
\def\bea{\begin{eqnarray}}
\def\eea{\end{eqnarray}}
\def\la{\langle}
\def\ra{\rangle}
\def\ef{{\rm erf}}
\begin{document}
\title{Weak values and the Leggett-Garg inequality in solid-state qubits}
\author{Nathan S. Williams and Andrew N. Jordan}
\affiliation{Department of Physics and Astronomy, University of Rochester, Rochester, New York 14627, USA}
\date{\today}
\begin{abstract}
An implementation of weak values is investigated in solid-state qubits. We demonstrate that a weak value can be non-classical if and only if a Leggett-Garg inequality can also be violated.  Generalized weak values are described, where post-selection on a range of weak measurement results.  Imposing classical weak values permits the derivation of  Leggett-Garg inequalities for bounded operators. Our analysis is presented in terms of kicked quantum nondemolition measurements on a quantum double-dot charge qubit.  
\end{abstract}
\pacs{73.23.-b,03.65.Ta,03.67.Lx}
\maketitle
The seminal paper of Aharonov, Albert, and Vaidman (AAV) introduces the concept of a {\it weak value} as a statistical  average over realizations of a weak measurement, where the system is both pre- and post-selected \cite{AAV}.
By taking restricted averages, weak values can exceed the range of eigenvalues associated with the observable in question  \cite{AAV,AB,Wiseman}. For example, AAV described how it would be possible to use a weak measurement to measure (say) the $\sigma_z$ eigenvalue of a spin-$1/2$ particle, and determine
an average value $\la\sigma_z\ra=100$.  This prediction of weak values that exceed the range of eigenvalues (hereafter referred to as {\it strange} weak values) has now been experimentally confirmed in quantum optics \cite{Wiseman Photon, First}, though there have been past \cite{Leggett} and ongoing \cite{controversy} debates as to the interpretation of this strange prediction.

In parallel activity, Leggett and Garg have devised a test of quantum mechanics for a single system using different ensembles of (projective) measurements at different times and correlation functions of those outcomes \cite{LG}. The original motivation was to test if there was a size scale where quantum mechanics would break down.  Introduced as a ``Bell-inequality in time'', the assumptions of Macrorealism (MAR) that could be verified by a Non-Invasive Detector (NID) imply that their correlation function obeys a Leggett-Garg inequality (LGI) that quantum mechanics would violate, formally similar to the inequality of Bell \cite{bell}.  This inequality has recently been generalized to weak measurements, using continuous \cite{ruskov} or discrete \cite{JKB} time correlation functions without the need for ensemble averaging over multiple configurations. 

The purpose of this paper is to demonstrate that a proper notion of the classicality of a weak value also requires the assumptions of MAR \& NID.  This fact also shows that a strange weak value can serve the same purpose envisioned by Leggett and Garg, namely as a test of macroscopic quantum coherence.  Furthering this connection, we demonstrate that a strange weak value (that requires averaging over a subset of post-selected data) can be observed if and only if a generalized LGI (that uses all the measurement data) can also be violated.

\begin{figure}[bh]
\begin{center}
\leavevmode \includegraphics[width=8cm]{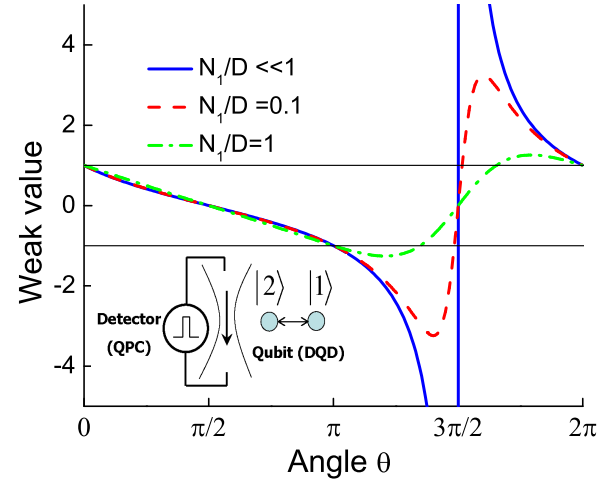} 
\caption{Post-selected weak value $_1\langle {\cal I}_1 \rangle_{\psi}$ starting from an initial state $|\psi\ra = (i|1\ra+|2\ra)/\sqrt{2}$ as a function of the unitary rotation angle between the measurements.  Strange weak values exceed the classical bound on the system signal, shown with horizontal lines at ${\cal I}=\pm 1$.  The weak value measurement is implemented by averaging the QPC detector current (${\cal I}_1$) over a subensemble where the later QPC current gives a particular value (here ${\cal I}_2=+1$).  The different curves show how the weak value changes for different measurement strengths.} 
\label{fig2}
\end{center}
\vspace{-5mm}
\end{figure}

Our results are discussed in terms of solid-state physics.   We consider a double quantum dot (DQD) quantum two-level system, with a capacitively coupled quantum point contact (QPC) detector ({\it c.f.} Fig. 1) \cite{expts}. Using stroboscopic ``kicked'' measurements, the position of the electron in the DQD is weakly measured.  Our results only depend on the ability to make weak measurements on qubits, and therefore also extend to {\it e.g.} macroscopic superconducting systems \cite{super}.  The weak value dependence on the strength of the measurement is presented, as well as a generalization where the post-selected averaging is over a range of weak measurement results.  Recent experimental advances in nano-scale semiconductor quantum dots have demonstrated (post-selected) single electron statistics from a non-invasive QPC detector, indicating that an experimental realization of weak values in the solid-state is feasible in the near future \cite{electron stats}.

{\it Measurement scheme and implementation.}---The DQD charge qubit is formed by quantum tunneling (with tunnel coupling energy $\Delta$) that hybridizes the ground states of two coupled quantum dots. The isolated DQD is described by qubit Hamiltonian $H_0 = (\epsilon \sigma_z + \Delta \sigma_x)/2$, where the energy asymmetry between the levels ($\epsilon$) is set to zero.  In order to measure a weak value it is necessary to make measurements of varying strength on this qubit. A convenient way of implementing this requirement for a qubit is to make weak, kicked measurements on the DQD charge operator ($\sigma_z$). The kicked measurements are separated in time by the qubit period, $\tau_{q} = 2\pi/\Delta$, and can be implemented with rapid periodic voltage pulses across the QPC and recording the current from each kick.  This quantum nondemolition (QND) measurement effectively eliminates the qubit Hamiltonian by going into the (stroboscopic) rotating frame \cite{JKB,JBJK}.   A second requirement for measuring a weak value is to apply controlled unitary operations to the qubit.  This is implemented in the kicked measurement scheme by inserting a ``dislocation'' in the pulse sequence by waiting a non-integer fraction $r$ of $\tau_{q}$ between successive kicks, $t_{\rm wait} = r \tau_q$, that defines a phase shift $\theta = 2 \pi r$.  

To characterize the result of each measurement kick, the parameters of the measurement process with an ideal QPC detector are specified by the currents, $I_1$ and $I_2$, produced by the detector when the qubit is in state $\vert 1\ra$ or $\vert 2\ra$ (see Fig.~1), and the detector shot noise power $S_I =e I (1-T)$ (where $T$ is the transparency) \cite{noteconv}. The typical integration time needed to distinguish the qubit signal from the background noise is the measurement time $T_M = 4S_I/(I_1-I_2)^2$.  Shifted, dimensionless variables may be introduced by defining the current origin at $I_0=(I_1+I_2)/2$, and scaling the current per pulse as $I -I_0= x (I_1-I_2)/2$, so $I_{1,2}$ are mapped onto $x=\pm 1$.  We take $x$ to be normally distributed with variance $D = T_M/\tau_V$.  The typical number of kicks needed to distinguish the two states is $D$, where we assume $D\gg 1$.  The dimensionless current $\cal I$ after $N$ kicks is ${\cal I} = (1/N)\sum_{n=1}^{N} x_n$.  

{\it Implications of a strange weak value.}---In order to have a precise meaning of the ``quantumness'' of the weak value, we first will examine what assumptions are necessary to have a classical weak value. 
Starting with a pre-selected state $|\psi\ra$, we first consider a weak measurement of variable strength ($N_1 \sim D$), separated by a phase shift $\theta$ from a second projective measurement ($N_2 \gg D$).  The assumption of MAR is that the measured system always has a well defined value that furthermore can be determined by a NID that does not alter the system.  MAR implies that the result ${\cal I}_n$ for the $n^{\rm th}$ measurement (which is generally composed of many QND kicks) can be decomposed as ${\cal I}_n = C_n + \xi_n$, 
where $C_n$ is the signal from the system (recall $-1 \le C_n \le 1$) and $\xi_n$ is a white noise source that describes the Gaussian shot noise of the detector \cite{noteMAR}.  This noise source satisfies 
\begin{math}
\langle \xi_n \rangle=0
\end{math} and 
\begin{math}
\langle \xi_n \xi_m \rangle=(D/N_n) \delta_{m,n},
\end{math}
where $D/N_n$ is width of the current distribution after $N_n$ kicks have been performed.  

We now consider the restricted average of the first weak measurement result, ${\cal I}_1$, post-selected on the results of the second projective measurement, ${\cal I}_2$.  We introduce the mixed notation $_{{\cal I}_2}\langle {\cal I}_1 \rangle_{\psi}$ for this post-selected weak value \cite{note}.   The linearity of (post-selected) averaging implies
\be
_{{\cal I}_2}\langle {\cal I}_1 \rangle_{\psi} = \,_{{\cal I}_2}\langle C_1 \rangle_{\psi} + \,_{{\cal I}_2}\langle \xi_1 \rangle_{\psi}.
\label{wv}
\ee
The second measurement is projective, so its uncertainty, $\langle \xi_2^2 \rangle = D/N_2$, vanishes as
$N_2 \rightarrow \infty$, leaving just the signal (${\cal I}_2 = C_2$) which can be $1$ or $-1$.  The first term in Eq.~(\ref{wv}) is just the post-selected  classical average of the signal, which can have any time dependence so long as it is bounded between $[-1, 1]$.  This term is therefore trivially bounded, \begin{math}-1 \le \,_{{\cal I}_2}\langle C_1 \rangle_{\psi} \le 1\end{math}.  The second term in Eq.~(\ref{wv}) can be analyzed by invoking NID:  the bare detector noise in the past ($\xi_1$) does not affect the system signal in the future (${\cal I}_2 = C_2$), so $\xi_1$ is uncorrelated with $C_2$.  For an uncorrelated variable the conditional (post-selected) and unconditional averages coincide, so $_{{\cal I}_2}\langle \xi_1 \rangle_{\psi_1} = \langle \xi_1 \rangle_{\psi_1} = 0$.  
Summing up, MAR \& NID imply 
\begin{equation}
-1 \le \,_{{\cal I}_2}\langle {\cal I}_1 \rangle_{\psi} \le 1.
\label{Weak Value Inequality}
\end{equation}
We stress the surprising result that NID must be invoked to have a bounded weak value.  Even for a classical system, if the detector is invasive, it will be possible for the post-selected weak value to exceed the signal bound of the system.

{\it Quantum Analysis.}---
We now reconsider the preceding situation from a quantum point of view.  The initial density matrix elements are $\rho_{ij}$ in the $\{ |1\ra, |2\ra\}$ basis.  The first weak measurement will give a specific QPC current, ${\cal I}_1$, which according to the quantum Bayesian approach \cite{sasha1,JKB} will be selected from the probability distribution
\begin{equation}
P({\cal I},N)= \rho_{11}P({\cal I},N|1)+\rho_{22}P({\cal I},N|2),
\label{prob dist}
\end{equation}
where $P({\cal I},N|x)$ is a Gaussian distribution with an average of $(-1)^{1+x}$, and a variance of $D/N$.  The first measurement alters the state of the qubit depending on the outcome of the measurement, ${\cal I}_1$, partially collapsing the state.  The new density matrix after the measurement is
\begin{equation}
\rho' = \frac{1}{\rho_{11}e^{\gamma_1}+\rho_{22}e^{-\gamma_1}}\begin{pmatrix} \rho_{11}e^{\gamma_1} & \rho_{12} \\ \rho_{12}^* & \rho_{22}e^{-\gamma_1} \end{pmatrix},
\end{equation}
where we define $\gamma_i = {\cal I}_i N_i/D$.  We now wait a non-integer multiple of the qubit period, $t_{\rm wait}$ before the next measurement kick.  This will cause a unitary rotation ${\bf U}_x(\theta)$ about the $x$-axis by an angle $\theta$, giving ${\tilde \rho} = {\bf U}_x\rho' {\bf U}_x^\dagger$ as the new density matrix.
The second projective measurement on the system is implemented with $N_2 \gg 1$ QND kicks, giving ${\cal I}_2 = 1$ with probability ${\tilde \rho}_{11}$ and ${\cal I}_2 = -1$ with probability ${\tilde \rho}_{22}$.  

Now the post-selection can be performed and the weak value calculated.  We have available the probability of measuring ${\cal I}_2$ given a specific ${\cal I}_1$ (either ${\tilde \rho}_{11}$ or ${\tilde \rho}_{22}$), but to average over the post-selected ensemble, we need the probability of measuring ${\cal I}_1$ given the chosen result ${\cal I}_2$.  Bayes' Theorem,
$P({\cal I}_1|{\cal I}_2) = P({\cal I}_2|{\cal I}_1)\, P({\cal I}_1) / P({\cal I}_2)$ allows us to calculate this conditional probability.  The weak values are then given by
\begin{equation}
_{{\cal I}_2}\langle {\cal I}_1 \rangle_{\psi} = \int\limits_{-\infty}^\infty {\cal I}_1 P({\cal I}_1 | {\cal I}_2) d{\cal I}_1.
\label{wv integral}
\end{equation}
Applying the result \eqref{prob dist} with $\rho \rightarrow {\tilde \rho}$, together with the new density matrix $\tilde \rho$, we find for the weak value \eqref{wv integral}
\begin{equation}
\begin{gathered}
_1\langle {\cal I}_1 \rangle_{\psi} = \frac{\cos^2(\frac{\theta}{2})\rho_{11} - \sin^2(\frac{\theta}{2})\rho_{22}}{\cos^2(\frac{\theta}{2})\rho_{11} + \sin^2(\frac{\theta}{2})\rho_{22} - \sin\theta\,{\rm Im}\rho_{12}e^{-S_1}},
\\
_{-1}\langle {\cal I}_1 \rangle_{\psi} = \frac{\sin^2(\frac{\theta}{2})\rho_{11} - \cos^2(\frac{\theta}{2})\rho_{22}}{\sin^2(\frac{\theta}{2})\rho_{11} + \cos^2(\frac{\theta}{2})\rho_{22} + \sin\theta\,{\rm Im}\rho_{12}e^{-S_1}},
\label{wv1}
\end{gathered}
\end{equation}
where $S_{1,2} = N_{1,2}/(2 D)$ is the {\it strength} of the measurement.  We will see shortly that the weak values can exceed the classical range [-1,1].
For an initial state $|\psi \rangle = (i|1\rangle + |2\rangle)/\sqrt{2}$, the weak values simplify to
\begin{equation}
_{\pm 1}\langle {\cal I}_1 \rangle_{\psi} = \frac{\pm \cos\theta}{1 \pm \sin\theta\,\exp(-S_1)}.
\label{wvspecial}
\end{equation}
In the limit $S_1 \rightarrow 0$ the weak values diverge for certain values of $\theta$.  For realistic, finite strength, measurements ($S_1 \sim 1$) there can still be intervals of $\theta$ that produce strange weak values (see Fig.~1).  Indeed, if we adjust the angle $\theta$ appropriately, it is possible to observe a strange weak value for {\it any} finite measurement strength.  Maximizing the ${\cal I}_2=1$ weak value (\ref{wvspecial}) as a function of $\theta$ (so $\theta_{\rm max} = - \arcsin \exp(-S_1)$), gives a maximal weak value
\begin{equation}
_1\langle {\cal I}_1 \rangle_{\psi}^{\rm max} = 1/\sqrt{1-\exp(- 2 S_1)},
\label{max}
\end{equation}
which exceeds $1$ for any finite strength $S_1$.

{\it Connection to the Leggett-Garg Inequality.}---
We now show that strange weak values are essentially equivalent to a violation of a generalized LGI.   The original LGI inequality was generalized to weak measurements by considering an experiment similar to the one described above, but with three weak measurements, ${\cal I}_A$, ${\cal I}_B$, and ${\cal I}_C$, of arbitrary strength, separated by two ${\bf U}_x(\theta)$ rotations with angles $\theta_1$ and $\theta_2$ \cite{JKB}.  The correlation function, $B=K_{AB}+K_{BC}-K_{AC}$ is defined, where $K_{nm} = \langle {\cal I}_n {\cal I}_m \rangle$ is the (unconditional) correlation function of the current results and $n,m = A,B,C$.  A quantum analysis showed that for any initial state,
\begin{equation}
B=\cos\theta_1 + \cos\theta_2 - \cos\theta_1\cos\theta_2 + \sin\theta_1\sin\theta_2 \exp(-S_B),
\label{JKB}
\end{equation}
while under the assumptions of MAR \& NID, this function is bounded, $-3\le B \le1 $.  With the choice of certain parameters (\ref{JKB}) can violate this inequality (maximal violation is for $S_B\ll1$, and $\theta_1 = \theta_2 = \pi/3$, so $B=3/2$).

We now reformulate a specific LGI in terms of weak values.  It is important to note that Eq.~(\ref{JKB}) has no dependence on the strength of the first or third measurement, so we consider the special case $N_1, N_3 \rightarrow \infty$ ({\it i.e.} the first and last measurements are projective).  For definiteness take $\psi_0=(1,0)$, so that the first measurement result deterministically gives ${\cal I}_A=1$ and ${\bf U}_x(\theta_1)$ rotates the initial state to $\psi=(\cos\theta_1, -i\sin\theta_1)$.  This procedure essentially uses the first projective measurement and subsequent unitary operation to prepare an initial state $\psi$ for the weak value measurement (we now identify ${\cal I}_B \rightarrow {\cal I}_1$ and ${\cal I}_C \rightarrow {\cal I}_2$ from the preceding analysis).  The correlation function now becomes
\begin{equation}
B = \langle {\cal I}_B \rangle_{\psi} + \langle {\cal I}_B {\cal I}_C \rangle_{\psi} - \langle {\cal I}_C \rangle_{\psi}.
\label{Bwv}
\end{equation}
Here we see that the generalized LGI needs only two measurements, together with their averages and correlation function. The last (projective) measurement (${\cal I}_C$) will take the value of 1 or -1.  This allows us to rewrite Eq.~(\ref{Bwv}) as
\begin{equation}
B = (2\ _{1}\langle {\cal I}_B \rangle_{\psi} - 1 )P_C(1) + P_C(-1),
\label{simplified B}
\end{equation}
where $_{1}\langle {\cal I}_B \rangle_{\psi}$ is the weak value post-selected on ${\cal I}_C=1$, and $P_C(\pm1)$ is the unconditional probability of measuring ${\cal I}_C = \pm 1$.  Noting $P_C(-1) = 1-P_C(1)$, the classical bound on the weak value (\ref{Weak Value Inequality}) then implies
\begin{equation}
-3 \le -4 P_C(1) + 1 = B \le P_C(1) + P_C(-1) = 1,
\label{endwv}
\end{equation}
which is the same inequality derived in Ref.~\cite{JKB} (though for a more specific case).  Thus, if a strange weak value is experimentally observed, then a generalized Leggett-Garg inequality will also be violated.

Furthermore the violation of this generalized LGI ($-3 \le B \le 1$) implies the existence of a strange weak value.  Recalling the equality (\ref{simplified B}) (again for the special case described above) we can convert the LGI into a bound on the weak value,
\begin{equation}
\begin{split}
-3 \le (2\ _1\langle {\cal I}_B \rangle_{\psi} - 1 )P_C(1) + 1 - P_C(1) &\le 1, \\
-4 \le (2\ _1\langle {\cal I}_B \rangle_{\psi} - 2) P_C(1) &\le 0, \\
-1 \le {_1\langle {\cal I}_B \rangle_{\psi}} &\le 1,
\end{split}
\end{equation}
unless $P_C(1)=0$ \cite{note2}.  Therefore, the generalized LGI discussed above will be violated if and only if the weak value is strange.  This is an interesting result since the generalized LGI {\it per se} requires no post-selection, only simple correlation functions \cite{note3}.

\begin{figure}[tbh]
\begin{center}
\leavevmode \includegraphics[width=8 cm]{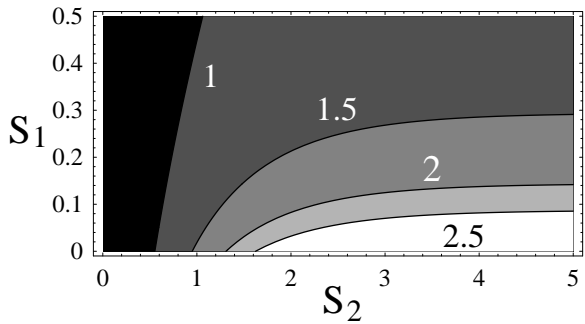} 
\caption{Post-selected weak value $_{[0,\infty)}\langle {\cal I}_1 \rangle_{\psi}^{\rm max}$ starting from an initial state $|\psi\ra = (i|1\ra+|2\ra)/\sqrt{2}$ as a function of the strength ($S_1$ and $S_2$) of both measurements.  Strange weak values exceed the classical bound on the system signal, and are outside the black region.  Weak values are labeled at the contour lines.} 
\label{fig3}
\end{center}
\vspace{-5mm}
\end{figure}

{\it Weak Values from a range of measurements.}---The most important practical limitation on the above implementation of weak values is the use of projective measurements; a limitation that the generalized LGI of Ref.~\cite{ruskov,JKB} does not have.   We now generalize the weak values of \cite{AAV} to the situation where the final measurement is weak (see also \cite{howard}), but post-selection occurs over a partition of the measurement results into ranges of the form ${\cal I}_2 \in [l,u]$.  Following the preceding analysis we define the generalized weak value
$_{[l,u]}\langle {\cal I}_2 \rangle_{\psi} = \int {\cal I}_1 P({\cal I}_1|{\cal I}_2\!\in\![l,u]) d{\cal I}_1$. This value can be calculated explicitly within our model in terms of error functions, but it is too lengthy to present here.  For definiteness, we again start with the state $|\psi \rangle = (i|1\rangle + |2\rangle)/\sqrt{2}$ and consider the specific post-selection range $[0,\infty)$, giving
\begin{equation}
_{[0,\infty)}\langle {\cal I}_1 \rangle_{\psi} = \frac{\cos\theta\, \ef [\sqrt{S_2}\, ]}{1 + \sin\theta\,\exp(-S_1)\,\ef[\sqrt{S_2}]}.
\label{wvrange}
\end{equation}
In Fig.~2, we plot the maximized weak value versus the measurement strengths $S_1$ and $S_2$.
This result demonstrates that the generalized strange weak value needs a much stronger second measurement than the first (unlike the LGI).  Contours lines label weak values; the strange ones are outside the black region.

{\it Generalizations.}---We briefly discuss the use of weak values to derive a LGI for operators $O$ with a finite number of (ordered) discrete eigenvalues, $\lambda_l\!<\!\ldots\!<\!\lambda_m$.  The weak value corresponding to this operator is $_f\la O\ra_i=\la \psi_f \vert O \vert \psi_i\ra/\la \psi_f|\psi_i\ra$ where $|\psi_{i,f}\ra$ is the (pre-)post-selected state \cite{AAV}.  We define the generalized LGI ${\tilde B} = \la O_A O_B\ra -  \la O_A O_C\ra + \la O_B O_C\ra$, where now the angle brackets denote quantum expectations of the measurements on the same operator $O$ at times $A$ and $C$ (which are projective), and time $B$ (which is arbitrarily weak in the AAV sense).  Following steps similar to (\ref{Bwv},{\ref{simplified B},\ref{endwv}), we can derive a LGI by imposing the eigenvalue bounds on the weak values, $\lambda_l\le {_f\la} O\ra_i \le \lambda_m$ to find
\be
{\tilde B} \le {\rm max}(\lambda_l^2, \lambda_m^2),
\ee
which will be violated by quantum mechanics (the lower bounds are cumbersome and will be presented elsewhere).  

{\it Conclusions.}---We have shown that any experiment that demonstrates a strange weak value must also violate a generalized LGI (notwithstanding \cite{Leggett}), though it is possible to violate other LGIs that are not amenable to a standard weak values analysis.  The LGI involves only simple correlation functions of the full data set, while weak values need post-selection (post-selection may, however, be considered an exotic correlation function \cite{howard}).  We have also investigated the conditions under which a weak value with a weak post-selection measurement can be strange. The ability to be given a single data record and say whether it was produced by a classical or quantum system is surprising, and provides a test of quantum coherence.  The continuing development of quantum-limited measurement in nanoelectronics should make these predictions realizable in the near future.

We thank Howard Wiseman for helpful discussions, and for suggesting a connection between the LGI and weak values.


\begin{thebibliography}{99}
\bibitem{noteconv}
Our convention for the spectral density is $\la I^2\ra -\la I
\ra^2 = \int_{-\infty}^{\infty} S_I(\omega) d\omega/(2\pi)$.
\bibitem{noteMAR}
We note MAR, as philosophy, is not limited to macroscopic systems only.
\bibitem{note}
We use a nonstandard notation here (post-selecting on the measurement result rather than the final state) because we will later post-select on a range of (weak) measurement results.
\bibitem{note2}
The original weak value \cite{AAV} is undefined at this point, so it does not compromise our result.
\bibitem{note3}
Starting with $B$ selects the ${\cal I}_C = 1$ weak value (\ref{simplified B}).  To select the ${\cal I}_C = -1$  weak value in the same way, we define another generalized LGI, $B'=K_{AB}-K_{BC}+K_{AC}$.
\end{thebibliography}
\end{document}